\documentclass[a4paper]{article}
\pdfoutput=1
\usepackage[english]{babel}
\usepackage[utf8]{inputenc}
\usepackage[T1]{fontenc}
\usepackage{authblk}
\usepackage{siunitx}
\usepackage[a4paper,top=3cm,bottom=2cm,left=3cm,right=3cm,marginparwidth=1.75cm]{geometry}
\usepackage{amsmath}
\usepackage{graphicx}
\usepackage[colorinlistoftodos]{todonotes}
\usepackage[colorlinks=true, allcolors=black,pdftex]{hyperref}

\title{Performance of a cryo-cooled crystal monochromator illuminated by hard X-rays with MHz repetition rate at the European X-ray Free-Electron Laser}
\author[1,*]{I.~Petrov}
\author[1]{U.~Boesenberg}
\author[2]{V.~A.~Bushuev}
\author[1]{J.~Hallmann}
\author[1]{K.~Kazarian}
\author[1]{W.~Lu}
\author[1]{J.~M\"oller}
\author[1]{M.~Reiser}
\author[1]{A.~Rodriguez-Fernandez}
\author[1]{L.~Samoylova}
\author[1]{M.~Scholz}
\author[1]{H.~Sinn}
\author[1]{A.~Zozulya}
\author[1]{A.~Madsen}
\affil[1]{European X-Ray Free-Electron Laser Facility, Holzkoppel 4, D-22869 Schenefeld, Germany}
\affil[2]{Lomonosov Moscow State University, 119992 GSP-2, Moscow, Russia}
\affil[*]{Email for correspondence: ilia.petrov@xfel.eu}
\date{}
\begin{document}
\maketitle
\begin{abstract}
Due to the high intensity and MHz repetition rate of photon pulses generated by the European X-ray Free-Electron Laser, the heat load on silicon crystal monochromators can become large and prevent ideal transmission in Bragg diffraction geometry due to crystal deformation. Here, we present experimental data illustrating how heat load affects the performance of a cryogenically cooled monochromator under such conditions. The measurements are in good agreement with a depth-uniform model of X-ray dynamical diffraction taking beam absorption and heat deformation of the crystals into account.
\end{abstract}

\section{Introduction}
Monochromators are often used at X-ray Free-Electron Lasers (XFELs) to reduce the spectral bandwidth of pulses generated by self-amplified spontaneous emission (SASE, bandwidth $\Delta E/E \sim 2\cdot10^{-3}$). This improves the temporal coherence, which is beneficial for several techniques. For instance, X-ray Photon Correlation Spectroscopy (XPCS) in Wide-Angle X-ray Scattering (WAXS) provides a better contrast when a monochromator is used \cite{Madsen2016,Lehmkuhlerro5014}. Also, high-resolution and inelastic X-ray scattering require narrower bandwidth than SASE provides \cite{ShvydkoIXS,Chubarixs}. Obtaining nano-sized foci with refractive optics (chromatic focusing) also benefits from a reduction of the X-ray bandwidth \cite{nanofocusmono}.

At European XFEL (EuXFEL), X-ray pulses with the intensity of several mJ and a few femtosecond duration are generated. The pulses are delivered in so-called "trains" which can contain several hundreds of pulses that are separated by  sub-\SI{600}{\micro\second} intervals. That is, within a train of typically $\sim$\SI{600}{\micro\second} duration, pulses arrive at MHz repetition rate and 10~trains are delivered per second. For the moment, 2.25~MHz repetition rate is available on a standard basis at EuXFEL but the design value of 4.5~MHz has also been achieved \cite{decking2020mhz}. 

The intense radiation causes deformation of the crystal lattice, which affects the diffraction of X-rays and degrades the monochromator performance. We study the performance of a cryogenically cooled Si(111) monochromator that consists of two parallel crystals forming an artificial channel-cut \cite{dongmono,petrovfel2019} (Darwin width $\Delta E/E\sim1.4\cdot10^{-4}$) operating at 9~keV photon energy using experimental data obtained at the Materials Imaging and Dynamics (MID) instrument of EuXFEL \cite{XFELbeamtrans,Madsenay5570}. The geometry of Si crystals (60~mm length, 35~deg. angular range of Bragg-angle rotation) enables to use the monochromator in a 5-25~keV photon energy range. To achieve parallel position of both Si crystals the second crystal can be adjusted in pitch and roll angles. Fine pitch adjustment with sub-microradian precision is enabled using a piezo-actuator. Both crystals are supplied with cryogenic cooling using a helium cryocompressor and equipped with heaters to achieve stable thermal conditions \cite{Sinn139077}. The temperature is typically set to ~100~K which is close to the zero point of the thermal expansion coefficient of silicon \cite{Lyonsilinexp}.

In the following, simulations which consider heat propagation during a train \cite{SinnHeatload} and dynamical diffraction effects \cite{Bushuev2013,Bushuev2016} (Sections \ref{HeatingSim}, \ref{DynDifrSim}) are compared with experimental data (Section \ref{Experimental}).

\section{Simulation of crystal heating}\label{HeatingSim}

For the simulation of heat absorption and transfer, the crystal is divided into $n$ cylindrical shells with inner radii of $r_i=(i-1)\cdot dr$. $i$ varies from $1$ to $n$ and the outer radii of the ith shell is $r_i+dr$ where $dr$ is the thickness of each cylindrical shell \cite{SinnHeatload}. Along the surface normal direction cylinders are divided into $m$ layers of $dz$ thickness and the position of the jth layer in the depth coordinate is $z_j=j\cdot dz$, where $j$ varies from $1$ to $m$. 

Let us consider a Gaussian pulse whose radial intensity profile reads

\begin{equation}
    \label{gaus_beam}
    I(r_i,z_j)=I_0S(r_i)\frac{\exp(-r_i^2/2\sigma^2)\exp(-z_j/a)}{2\pi \sigma^2a}dz,
\end{equation}
where $I_0$ is the total pulse energy, $S(r_{i>1})=2\pi r_i dr$, $S(r_1)=\pi\cdot dr^2$, $\sigma=w_{\text{equiv}}/2\sqrt{2\ln{2}}$ where $w_\text{equiv}$ is the full-width at half maximum (FWHM) of the beam size at the crystal surface and $a$ is the depth at which the intensity of the beam decreases by a factor of $e$. Since the X-rays impinge the crystal at an angle $\theta_0$, the pulse size $w_\text{equiv}=w/\sqrt{\sin\theta_0}$ is used for the simulations, where $w$ is the FWHM size of the pulse incident at an angle $\theta_0$. $a=l_\text{abs}\sin{\theta_0}$, where $l_\text{abs}$ is the absorption length of X-rays at a given photon energy. The heat load per unit of surface area for the pulse with the size $w_\text{equiv}$ is the same as of the pulse with the size $w$ incident at an angle $\theta_0$.

The temperature of each cylinder layer with inner radius $r_i$ at depth $z_j$ is determined by the heat absorbed per unit of mass. The absorption of an incident pulse and resulting heating are considered to be instantaneous in comparison with the characteristic time for the redistribution of temperature (see below for the estimations of the timescales using Eq.~\ref{estim_time}). The temperature $T_0(r,z)$ at each radius $r$ and depth $z$ (indices of $r_i$ and $z_j$ are omitted) after the absorption of a pulse is defined by the absorbed heat per unit of mass in the corresponding cylindrical shell given by

\begin{equation}
    \label{Tinit}
    \int_{T_\text{init}}^{T_0(r,z)}c_p(T)dT=\frac{I(r,z)}{dz\cdot\rho S(r)},
\end{equation}
where $c_p(T)$ is the temperature-dependent specific heat of silicon that has been calculated following Debye's model\cite{debyespecheat}, $\rho$ is the density of silicon (whose temperature dependence is neglected), and $T_\text{init}$ is the initial temperature of the crystal. The temperature evolution with time $T(t,r,z)$ is defined by the heat transfer equation which in the depth direction is written as

\begin{equation}
    \label{heat_transf}
    \frac{\partial T(t,r,z)}{\partial t}=D(T)\cdot\frac{\partial^2 T(t,r,z)}{\partial z^2},
\end{equation}
where $D(T)=K(T)/\rho c_p(T)$ is the temperature-dependent thermal diffusivity and $K(T)$ is the temperature-dependent thermal conductivity \cite{lambdaSi}. The boundary conditions for Eq.~(\ref{heat_transf}) are

\begin{subequations}
\label{bound_heat}
\begin{align}
 T(0,r,z) = T_0(r,z),\\
\left.\frac{\partial T}{\partial z}\right|_{z=0}=0,\\
T(t,r,z=z_m)=T_\text{init},
\end{align}
\end{subequations}
which correspond to the absence of heat exchange at the crystal surface and a constant temperature $T_\text{init}$ at depth $z_m$. If a second pulse arrives at an instant $t_1$,  the temperature profile $T'(t_1,r,z)$ is defined analogous to Eq.~(\ref{Tinit}):

\begin{equation}
    \label{Tinit_1}
    \int_{T(t_1,r,z)}^{T'(t_1,r,z)}c_p(T)dT=\frac{I(r,z)}{\rho\cdot dz\cdot S(r)}.
\end{equation}

Let us analyze Eq.~(\ref{heat_transf}) in order to estimate the characteristic timescale of heat transfer in the radial and depth directions. The parameters used in the simulations match the parameters of the experiment described in Sec. \ref{Experimental}: $T_{\text{init}}=100$~K, FWHM beam size of \SI{549}{\micro\meter}, $a=$~\SI{21}{\micro\meter} corresponding to $l_\text{abs}=$~\SI{97}{\micro\meter} and $\theta_0=12.7${\textdegree}, which is the Bragg angle for Si(111) reflection at 9~keV photon energy. For silicon, $D(100~K)\approx17$~$\text{cm}^{-2}\text{s}^{-1}$. Considering the heat flow equation (\ref{heat_transf}), the characteristic time for the redistribution of temperature over a distance $L$ can be estimated as

\begin{equation}
    \label{estim_time}
    t_{\text{char}}(L)\sim\frac{L^2}{D}.
\end{equation}
Let us consider two relevant distances in Eq.~(\ref{estim_time}) $L_r=$~\SI{549}{\micro\meter}, which is equal to the beam size, and in the depth direction $L_z=a=$~\SI{21}{\micro\meter}. For the heat redistribution in the radial direction, $t_\text{char,r}=t_\text{char}(L_r)=$~\SI{181}{\micro\second}, whereas in the depth direction $t_\text{char,z}=t_\text{char}(L_z)=265$~ns. Hence $t_\text{char,r} >> t_\text{char,z}$.

The duration of pulses at European XFEL is estimated to be $\sim$10-100~fs, which is many orders of magnitude shorter than the characteristic heat redistribution time, see Eq.~(\ref{estim_time}). Therefore the assumption of instantaneous heating of the crystal by a pulse is justified. Moreover, the delay time between individual pulses at European XFEL is typically between 220 and 880~ns which is about two orders of magnitude smaller than $t_{\text{char,r}}$. Thus, neglecting heat flow in the radial direction is justified from one pulse to the next one inside the pulse train. However, $t_\text{char,z}$ is of the same order of magnitude as the time delay between pulses and therefore heat flow in the depth direction during a train must be accounted for in the simulations. On the other hand, the 0.1\,s interval between pulse trains is much larger than both $t_\text{char,r}$ and $t_\text{char,z}$ and therefore, by the time the next pulse train arrives, the crystal has fully recovered to the initial temperature and hence a non-deformed state.

In the experiment, the crystal is 2 cm thick and is kept at a constant cryogenic temperature. Therefore, the boundary condition Eq.~(\ref{bound_heat}c) defining a constant temperature at depth $z_m$ is applicable.

\section{Dynamical diffraction simulations}\label{DynDifrSim}
In the framework of kinematical diffraction, the resulting amplitude in a point with radius-vector $\vec{\varrho}$ is defined by the integral

\begin{equation}
    \label{kinem_diffr}
    E_s\propto\int\displaylimits_{\varrho_0}\frac{\exp(ik|\vec{\varrho}-\vec{\varrho_0}|)}{|\vec{\varrho}-\vec{\varrho_0}|}d^3\varrho_0,
\end{equation}
which represents the sum of waves scattered by each point of a scattering object with radius-vectors $\vec{\varrho_0}$ and where $k=2\pi/\lambda$ with $\lambda$ being the wavelength. In case of a crystal lattice where the atoms are positioned in a regular manner, secondary scattering of X-rays will have a significant effect on the resulting diffraction amplitude. This re-scattering in ideal crystals is described by the theory of dynamical diffraction of X-rays \cite{authier}. The Bragg law

\begin{equation}
    \label{Bragg}
    2d\sin\theta_B=\lambda
\end{equation}
represents the condition for coherent addition of waves scattered by a lattice and defines the Bragg angle $\theta_B$ at which the strongest scattering is observed for the given lattice spacing $d$.

The beam induced heating of the crystal described in the previous section causes a deformation of the lattice, which is different in each point of the crystal. Considering dynamical diffraction, only the component of the deformation normal to the crystal surface is relevant, since this affects the lattice spacing used in Eq.~(\ref{Bragg}). In order to estimate the effect of crystal deformation on the diffraction, we consider the heating of the crystal on the surface, i.e. at $z=0$. The lattice deformation $\epsilon(t,r)$ in the direction normal to the crystal surface caused by heating from $T_\text{init}$ to $T(t,r,z)$ is given by the accumulated expansion and

\begin{equation}
    \label{deformation}
    \epsilon(t,r)\equiv\frac{\Delta d(t,r)}{d_\text{init}}=\int_{T_\text{init}}^{T(t,r,z=0)}\alpha_T\left(T\right) dT,
\end{equation}
where $\alpha_T(T)$ is the temperature-dependent linear expansion coefficient of silicon which is close to zero near 100~K \cite{silinexp}, $d_\text{init}$ is the lattice spacing at temperature $T_\text{init}$ and $\Delta d(t,r)$ is the lattice spacing change after heating from $T_\text{init}$.

We assume that the Bragg condition is fulfilled for a photon energy $E_0=hc/\lambda$ in the case of a non-deformed crystal lattice. The diffraction of X-rays with a photon energy $E$ at an instant $t$ and at radius $r$ is defined by the deviation of the wave vector from the exact Bragg condition \cite{Bushuev2013,Bushuev2016}

\begin{equation}
    \label{alphaDynTheory_vec}
    \eta(E,t,r)=\frac{k^2-(\vec{k}+\vec{h}(t,r))^2}{k^2},
\end{equation}
where $\vec{k}$ is the wave vector for the incident beam with photon energy $E$ and $\vec{h}(t,r)$ is the reciprocal lattice vector $h(t,r)=2\pi/d_\text{init}[1+\epsilon(t,r)]$. In the simulations we assume that all photon energies are incident at the same angle $\theta_B$ (the Bragg angle for photon energy $E_0$ of the non-deformed crystal) and hence Eq.~(\ref{alphaDynTheory_vec}) can be written as

\begin{equation}
    \label{alphaDynTheory}
    \eta\left(E,t,r\right)=2\sin{2\theta_\text{B}}\left(\frac{\Delta E}{E_0}+\epsilon(t,r)\right)\tan{\theta_\text{B}},
\end{equation}
where $\Delta E=E-E_0$.

The reflection amplitude from an infinitely thick crystal is calculated as \cite{BushuevJSR2008}

\begin{equation}
    \label{Refl}
    R(E,t,r)=\frac{\eta(E,t,r)\pm\sqrt{\eta(E,t,r)^2-4\chi_h^2}}{2\chi_h},|R|\leq1,
\end{equation}
where $\chi_h$ is the first Fourier component of the crystal susceptibility. Eq.~(\ref{Refl}) defines the reflection amplitude at each point of the crystal surface and at a given photon energy $E$. The total reflection intensity from the crystal is defined as an integral of Eq.~(\ref{Refl}) over the crystal surface

\begin{equation}
    \label{Integr_2cryst_RC}
    I_E\left(E,t\right)\sim
    \int\displaylimits_0^{r_n}|R_0\left(E,t,r\right)|^2\cdot |R\left(E,t,r\right)|^2 \cdot
    \exp\left(-\frac{r^2}{2\sigma^2}\right)\frac{r}{\sigma^2} dr,
\end{equation}
where $R_0\left(E,t,r\right)$ is the reflection amplitude (\ref{Refl}) for the non-deformed crystal, i.e. $\epsilon\left(E,t,r\right)\equiv 0$.

The spectral width of the Bragg reflection of Si(111) at 9~keV is $\sim$1~eV, whereas the spectral width of the XFEL pulses is $\sim$20~eV. Thus only a narrow fraction of X-rays is reflected by the first crystal of the monochromator. Therefore we assume that the second crystal remains unheated and thus non-deformed and oriented parallel to the first crystal. The reflectivity from two crystals $I_R\left(t\right)$ in that case can be calculated as an integral of the reflection intensity (\ref{Integr_2cryst_RC}) over the photon energies as follows:
\begin{equation}
    \label{Integr2bounceInt}
    I_R\left(t\right)\sim\int \displaylimits_{\Delta E_0}I_E(E,t)dE,
\end{equation}
where $\Delta E_0$ is the range of photon energies.

\begin{figure}[ht!]
\centering\includegraphics[width=0.9\textwidth]{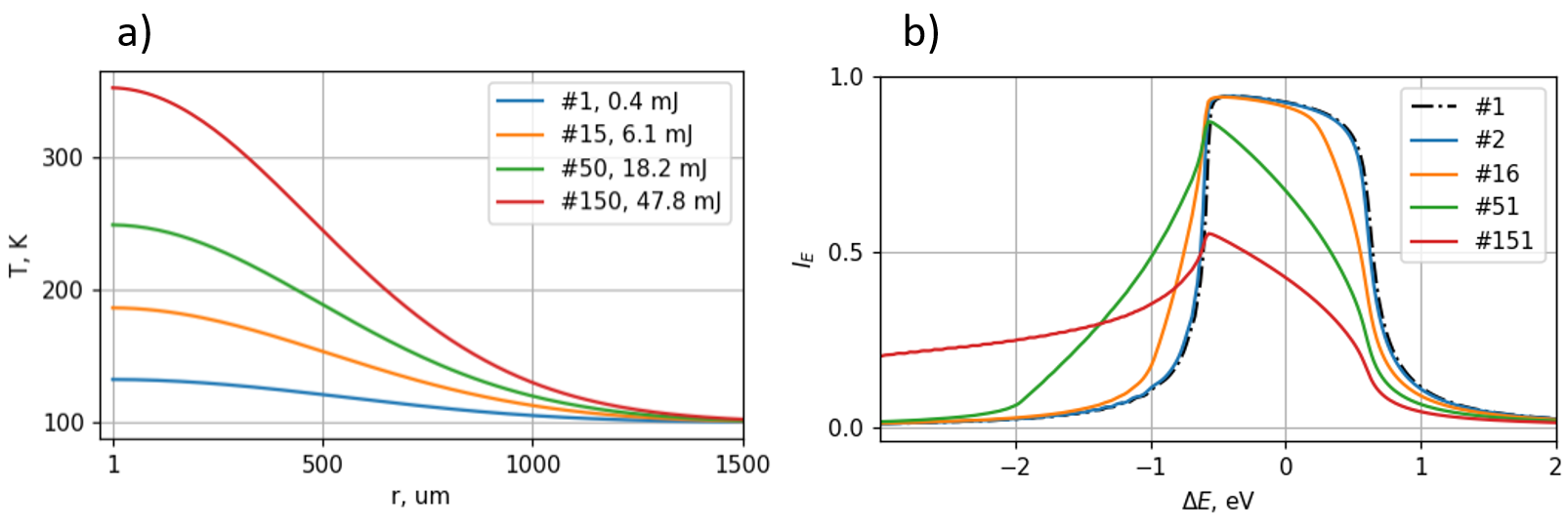}
\caption{Simulations for the effect of heating on the cryo-cooled Si monochromator performance. a) - simulated temperature profile at surface for various pulses in a train, the legend provides the total energy that has impinged the crystal since the beginning of the pulse train for a given pulse number, the energy of each pulse matches the ones measured at experiment, see inset in Fig.~\ref{transm_exp_sim}. b) - reflection intensity (\ref{Integr_2cryst_RC}) from the crystal within a range of photon energies for the temperature distributions in a).}
\label{simul_T_RC}
\end{figure}

Let us analyze the effect of heating on the monochromator performance using the parameters of the experiment described in Sec. \ref{Experimental}: pulse size $w=549$~\textmu m, repetition rate $2.25$~MHz, $T_\text{init}=100$~K, pulse energy ranging between 1 and $1.5$~mJ, as shown in the inset in Fig.~\ref{transm_exp_sim}, and the simulations are done for 30\% of the pulse energy impinging on the monochromator. The temperature in the center of the beam footprint reaches values in excess of 300~K, see Fig.~\ref{simul_T_RC}a), and Bragg's condition is no longer fulfilled because, due to the temperature bump, it is several Darwin widths away from the condition that was met at 100~K \cite{petrovfel2019}. Nevertheless, even when the temperature in the center of the illuminated area is high, there are areas of the crystal that are cold enough to reflect within the acceptance of the second crystal, which is assumed to stay cold at 100 K. The temperature gradient over the illuminated crystal area leads to a broadening of the reflectivity curve and a decrease of the reflected intensity in Fig.~\ref{simul_T_RC}b). At the end of the train of 150 pulses arriving at 2.25~MHz repetition rate, i.e. after a total of 47.8~mJ pulse energy has been absorbed by the first crystal, the monochromator transmission has decreased to less than half of the initial value.

\section{Experimental}\label{Experimental}
In order to measure the intensity of the pulses after the monochromator, a porous silica (Vycor) sample was used to scatter X-rays in the forward direction (small-angle X-ray scattering, SAXS).

\begin{figure}[ht!]
\centering\includegraphics[width=0.7\textwidth]{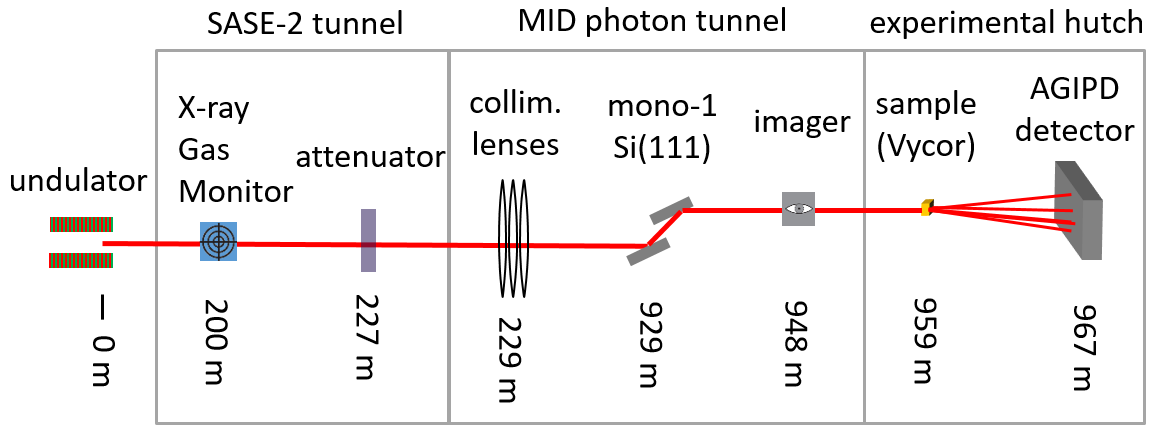}
\caption{Selected components of MID station at European XFEL and their positions relative to the source.}
\label{layout}
\end{figure}

\begin{figure}[ht!]
\centering\includegraphics[width=0.8\textwidth]{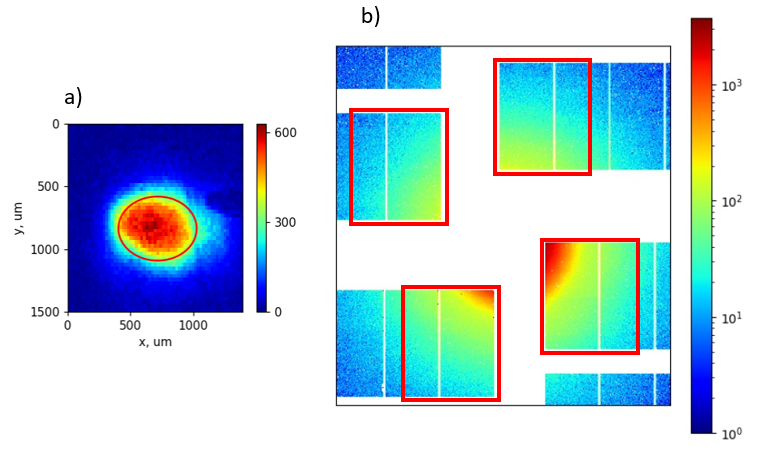}
\caption{a) intensity distribution of a pulse after monochromator on the YAG imager, the red ellipse is a contour at the FWHM of the two-dimensional Gaussian fit. b) Average over 60 images of the AGIPD area with the strongest SAXS signal. The red solid-line rectangles in b) denote the four areas of the detector used for data analysis.}
\label{1stframe_zoom}
\end{figure}

An overview of the beamline layout used at the experiment is shown in Fig.~\ref{layout}. The two-dimensional intensity distribution of the beam was measured using the yttrium aluminium garnet (YAG) screen imager device at the end of MID photon tunnel. The size of the beam was found by applying a two-dimensional Gaussian fit to the intensity distribution, see Fig.~\ref{1stframe_zoom}a). The horizontal FWHM width of the Gaussian fit $w_x=607$~\textmu m, the vertical - $w_y=496$~\textmu m; in the simulations, $w=\sqrt{w_xw_y}=549$~\textmu m, such that the average density of the circular pulse that is used in Eq.~(\ref{gaus_beam}) is equivalent to the elliptical beam shown in Fig.~\ref{1stframe_zoom}a). The scattered SAXS intensity was measured by the Adaptive Gain Integrating Pixel Detector (AGIPD) megapixel detector, which is designed to acquire full-frame data at frequencies up to 4.5 MHz \cite{AGIPDALLAHGHOLI2019162324}. The pulse intensity incident on the monochromator is measured using the X-ray gas monitor (XGM) device\cite{Maltezopoulosxt5015} installed after the undulator. Attenuators installed after the XGM are used to reduce the photon flux on the monochromator and the attenuator transmission was 30\% during the experiment. Collimating compound refractive lenses (CRLs) were used to compensate for the divergence of the beam\cite{rothcrl}.

In order to measure the scattering from the Vycor sample on AGIPD, only the pixels located closest to the center of the detector and having the strongest scattering signal were used for analysis of the monochromator transmission (Fig.~\ref{1stframe_zoom}b). The ratio of the sum of the intensity captured by the selected pixels to the XGM value provides a figure of merit for the transmission of a given pulse in a given train. Averaging of this ratio over a large number of trains for each pulse number provides an estimate of the monochromator transmission dependency on the energy that has impinged on the first crystal.

\begin{figure}[ht!]
\centering\includegraphics[width=0.8\textwidth]{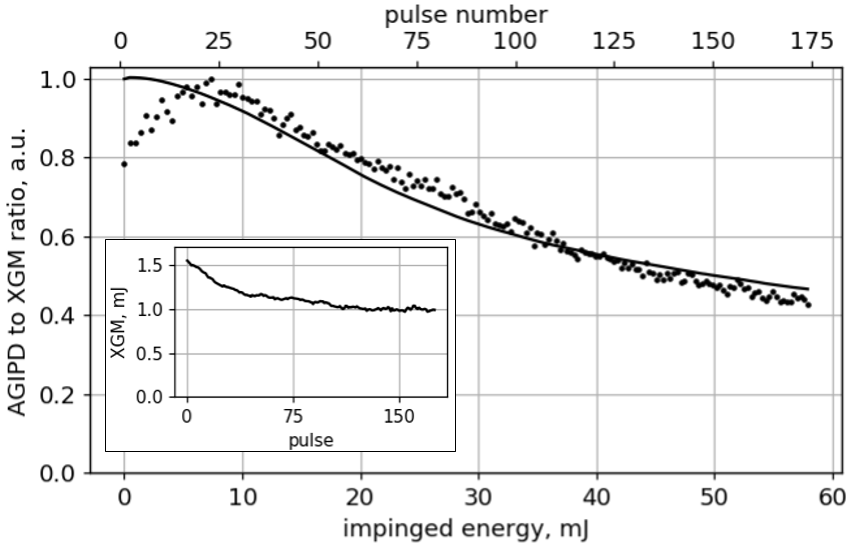}
\caption{Experimental (dots) and theoretical (line)  monochromator transmission during a train of XFEL pulses. The separation between pulses is 0.44~\textmu s, which corresponds to 2.25~MHz repetition rate. A photon energy 9~keV and a Si(111) reflection was used. The horizontal axis at the top represents the pulse number, at the bottom - the total energy that has impinged on monochromator before the respective pulse. The inset shows the energy of each pulse in a train measured by the XGM and averaged over the trains with 30\% of the energy impinges the monochromator. The experimental monochromator transmission is calculated as the ratio of the sum of AGIPD pixels to the XGM signal for each pulse in a train, averaged over 498 trains and normalized to the maximum value. The experimental and theoretical transmission values are normalized to the maximum values during the train.}
\label{transm_exp_sim}
\end{figure}

The measurements show that the monochromator transmission reduces by a factor of two after $\sim$50~mJ of X-ray energy or around 150 pulses under the aforementioned conditions, have impinged on the first crystal at 2.25~MHz repetition rate (Fig.~\ref{transm_exp_sim}). The experimental curves are not shown with error bars, since the transmission values are averaged over many trains. That is, for a fixed pulse number in a train, the scattering is produced by statistically independent and intrinsically random SASE pulses \cite{SALDIN1998383}. Even for an ideal monochromator the transmission is determined by the spectral intensity of the pulse in a bandwidth given by the Darwin width of the monochromator. Due to the random nature of the spectral fine structure of SASE\cite{SALDIN1998383}, averaging over a large number of pulses provides an accurate estimate of the effect of heating on the monochromator transmission. We attribute the initial rise of the measured monochromator transmission seen in Fig.~\ref{transm_exp_sim} to possible systematic drifts of photon energies and/or beam pointing during a pulse train.

The good agreement between theoretical and experimental values in Fig.~\ref{transm_exp_sim} indicates that the simulation model presented in Secs. \ref{HeatingSim} and \ref{DynDifrSim} provides a qualitatively correct behaviour of monochromator transmission during heating by intense X-ray pulses. Therefore the model can be employed as a simulation framework to aid the design of crystal optical devices when a high heat load from intense XFEL pulses is anticipated. The implementation of the code in Python is available to the public \cite{Petrovgithub}.

\section{Conclusion}
The intra-train transmission of a double-bounce Si(111) cryo-cooled monochromator has been measured at European XFEL using SASE pulses arriving at 2.25~MHz repetition rate. It has been shown that after around 150~pulses, which corresponded in this case to a total incident energy of around 50~mJ, the monochromator transmission decreases by about a factor of two.

A simple one-dimensional model of crystal heating and dynamical diffraction qualitatively reproduces the measured monochromator transmission. A simulation code is made available to the public \cite{Petrovgithub} and can be used to simulate the heat load effect on perfect crystal optical elements at XFELs.

\section*{Acknowledgments}
We are thankful to Roman Shayduk for fruitful discussions. We acknowledge EuXFEL for provision of beam time and would in particular like to thank the optics and vacuum groups for their help in designing and operating the cryo-cooled monochromator. The photon diagnostics group is acknowledged for enabling pulse-resolved X-ray pulse energy measurements using the XGM. The data department at EuXFEL is acknowledged for IT assistance, data acquisition, data analysis, and detector operation. AGIPD was developed and built by a consortium led by the photon science detector group at DESY.
\section*{Disclosures}
The authors declare no conflicts of interest.
\section*{Data availability} Data underlying the results presented in this paper are not publicly available at this time but may be obtained from the authors upon reasonable request.

\bibliographystyle{ieeetr}
\bibliography{mid_mono}

\begin{thebibliography}{10}

\bibitem{Madsen2016}
A.~Madsen, A.~Fluerasu, and B.~Ruta, {\em Structural Dynamics of Materials
  Probed by X-Ray Photon Correlation Spectroscopy}, pp.~1617--1641.
\newblock Cham: Springer International Publishing, 2016.

\bibitem{Lehmkuhlerro5014}
F.~Lehmk{\"{u}}hler, J.~Valerio, D.~Sheyfer, W.~Roseker, M.~A. Schroer,
  B.~Fischer, K.~Tono, M.~Yabashi, T.~Ishikawa, and G.~Gr{\"{u}}bel,
  ``{Dynamics of soft nanoparticle suspensions at hard X-ray FEL sources below
  the radiation-damage threshold},'' {\em IUCrJ}, vol.~5, pp.~801--807, Nov
  2018.

\bibitem{ShvydkoIXS}
Y.~Shvyd'ko, S.~Stoupin, D.~Shu, S.~Collins, K.~Mundboth, J.~Sutter, and
  M.~Tolkiehn, ``High-contrast sub-millivolt inelastic x-ray scattering for
  nano- and mesoscale science,'' {\em Nature communications}, vol.~5, p.~4219,
  06 2014.

\bibitem{Chubarixs}
O.~Chubar, G.~Geloni, V.~Kocharyan, A.~Madsen, E.~Saldin, S.~Serkez,
  Y.~Shvyd'ko, and J.~Sutter, ``{Ultra-high-resolution inelastic X-ray
  scattering at high-repetition-rate self-seeded X-ray free-electron lasers},''
  {\em Journal of Synchrotron Radiation}, vol.~23, pp.~410--424, Mar 2016.

\bibitem{nanofocusmono}
F.~Seiboth, A.~Schropp, R.~Hoppe, V.~Engemaier, J.~Patommel, H.~Lee, B.~Nagler,
  E.~Galtier, B.~Arnold, U.~Zastrau, J.~Hastings, D.~Nilsson, F.~Uhlén,
  U.~Vogt, H.~Hertz, and C.~Schroer, ``Focusing xfel sase pulses by
  rotationally parabolic refractive x-ray lenses,'' {\em Journal of Physics:
  Conference Series}, vol.~499, p.~012004, 04 2014.

\bibitem{decking2020mhz}
W.~Decking, S.~Abeghyan, P.~Abramian, A.~Abramsky, A.~Aguirre, C.~Albrecht,
  P.~Alou, M.~Altarelli, P.~Altmann, K.~Amyan, {\em et~al.}, ``{A
  MHz-repetition-rate hard X-ray free-electron laser driven by a
  superconducting linear accelerator},'' {\em Nature Photonics}, vol.~14,
  no.~6, pp.~391--397, 2020.

\bibitem{dongmono}
X.~Dong, D.~Shu, and H.~Sinn, ``{Design of a cryo-cooled artificial channel-cut
  crystal monochromator for the European XFEL},'' {\em AIP Conference
  Proceedings}, vol.~1741, no.~1, p.~040027, 2016.

\bibitem{petrovfel2019}
I.~Petrov, J.~Anton, U.~Boesenberg, M.~Dommach, X.~Dong, J.~Eidam, J.~Hallmann,
  K.~Kazarian, S.~Kearney, C.~Kim, W.~Lu, A.~Madsen, J.~Möller, M.~Reiser,
  L.~Samoylova, R.~Shayduk, D.~Shu, H.~Sinn, V.~Sleziona, and A.~Zozulya,
  ``{Effect of Heat Load on Cryo-Cooled Monochromators at the European X-Ray
  Free-Electron Laser: Simulations and First Experimental Observations},'' in
  {\em Proc. FEL'19}, pp.~502--505, nov 2019.

\bibitem{XFELbeamtrans}
T.~Tschentscher, C.~Bressler, J.~Grünert, A.~Madsen, A.~P. Mancuso, M.~Meyer,
  A.~Scherz, H.~Sinn, and U.~Zastrau, ``{Photon Beam Transport and Scientific
  Instruments at the European XFEL},'' {\em Applied Sciences}, vol.~7, no.~6,
  2017.

\bibitem{Madsenay5570}
A.~Madsen, J.~Hallmann, G.~Ansaldi, T.~Roth, W.~Lu, C.~Kim, U.~Boesenberg,
  A.~Zozulya, J.~M{\"{o}}ller, R.~Shayduk, M.~Scholz, A.~Bartmann, A.~Schmidt,
  I.~Lobato, K.~Sukharnikov, M.~Reiser, K.~Kazarian, and I.~Petrov,
  ``{Materials Imaging and Dynamics (MID) instrument at the European X-ray
  Free-Electron Laser Facility},'' {\em Journal of Synchrotron Radiation},
  vol.~28, pp.~637--649, Mar 2021.

\bibitem{Sinn139077}
H.~Sinn, M.~Dommach, X.~Dong, D.~La~Civita, L.~Samoylova, R.~Villanueva, and
  F.~Yang, ``{T}echnical {D}esign {R}eport: {X}-{R}ay {O}ptics and {B}eam
  {T}ransport,'' {\em Technical Report}, 2012.

\bibitem{Lyonsilinexp}
K.~G. Lyon, G.~L. Salinger, C.~A. Swenson, and G.~K. White, ``Linear thermal
  expansion measurements on silicon from 6 to 340 {K},'' {\em Journal of
  Applied Physics}, vol.~48, no.~3, pp.~865--868, 1977.

\bibitem{SinnHeatload}
H.~Sinn, ``Heat load estimates for {XFEL} beamline optics,'' {\em HASYLAB
  Annual Report 2007}, 2007.

\bibitem{Bushuev2013}
V.~A. Bushuev, ``Effect of the thermal heating of a crystal on the diffraction
  of pulses of a free-electron x-ray laser,'' {\em Bulletin of the Russian
  Academy of Sciences: Physics}, vol.~77, 01 2013.

\bibitem{Bushuev2016}
V.~A. Bushuev, ``Influence of thermal self-action on the diffraction of
  high-power x-ray pulses,'' {\em Journal of Surface Investigation. X-ray,
  Synchrotron and Neutron Techniques}, vol.~10, pp.~1179--1186, Nov 2016.

\bibitem{debyespecheat}
P.~Debye, ``Zur {Theorie} der spezifischen {Wärmen},'' {\em Annalen der
  Physik}, vol.~344, no.~14, pp.~789--839, 1912.

\bibitem{lambdaSi}
C.~J. Glassbrenner and G.~A. Slack, ``{Thermal Conductivity of Silicon and
  Germanium from 3${}^\circ$K to the Melting Point},'' {\em Phys. Rev.},
  vol.~134, pp.~A1058--A1069, May 1964.

\bibitem{authier}
A.~Authier, {\em Dynamical Theory of X-Ray Diffraction}.
\newblock Oxford, 2004.

\bibitem{silinexp}
Y.~Okada and Y.~Tokumaru, ``{Precise determination of lattice parameter and
  thermal expansion coefficient of silicon between 300 and 1500 K},'' {\em
  Journal of Applied Physics}, vol.~56, no.~2, pp.~314--320, 1984.

\bibitem{BushuevJSR2008}
V.~A. Bushuev, ``{Diffraction of X-ray free-electron laser femtosecond pulses
  on single crystals in the Bragg and Laue geometry},'' {\em Journal of
  Synchrotron Radiation}, vol.~15, pp.~495--505, Sep 2008.

\bibitem{AGIPDALLAHGHOLI2019162324}
A.~Allahgholi, J.~Becker, A.~Delfs, R.~Dinapoli, P.~Göttlicher, H.~Graafsma,
  D.~Greiffenberg, H.~Hirsemann, S.~Jack, A.~Klyuev, H.~Krüger, M.~Kuhn,
  T.~Laurus, A.~Marras, D.~Mezza, A.~Mozzanica, J.~Poehlsen, O.~{Shefer
  Shalev}, I.~Sheviakov, B.~Schmitt, J.~Schwandt, X.~Shi, S.~Smoljanin,
  U.~Trunk, J.~Zhang, and M.~Zimmer, ``Megapixels @ {Megahertz} – {The AGIPD}
  high-speed cameras for the {European XFEL},'' {\em Nuclear Instruments and
  Methods in Physics Research Section A: Accelerators, Spectrometers, Detectors
  and Associated Equipment}, vol.~942, p.~162324, 2019.

\bibitem{Maltezopoulosxt5015}
T.~Maltezopoulos, F.~Dietrich, W.~Freund, U.~F. Jastrow, A.~Koch, J.~Laksman,
  J.~Liu, M.~Planas, A.~A. Sorokin, K.~Tiedtke, and J.~Gr{\"{u}}nert,
  ``{Operation of X-ray gas monitors at the European XFEL},'' {\em Journal of
  Synchrotron Radiation}, vol.~26, pp.~1045--1051, Jul 2019.

\bibitem{rothcrl}
T.~Roth, L.~Helfen, J.~Hallmann, L.~Samoylova, P.~Kwaśniewski, B.~Lengeler,
  and A.~Madsen, ``{X-ray laminography and SAXS on beryllium grades and lenses
  and wavefront propagation through imperfect compound refractive lenses},'' in
  {\em Advances in X-Ray/EUV Optics and Components IX}, vol.~9207, pp.~1 -- 12,
  SPIE, 2014.

\bibitem{SALDIN1998383}
E.~Saldin, E.~Schneidmiller, and M.~Yurkov, ``{Statistical properties of
  radiation from VUV and X-ray free electron laser},'' {\em Optics
  Communications}, vol.~148, no.~4, pp.~383--403, 1998.

\bibitem{Petrovgithub}
\url{https://github.com/ia-petrov/cryst-heat-load}.

\end{thebibliography}

\end{document}